\documentclass[aps, reprint, floatfix, amsmath, amssymb, prl]{revtex4-2}
\usepackage{float} 
\usepackage{graphicx}
\usepackage{dcolumn}
\usepackage{bm}
\usepackage{xcolor}
\usepackage{xfrac}
\begin{document}

\title{Macro‑Dipole‑Constrained Learning of Atomic Charges \\ for Accurate Electrostatic Potentials at Electrochemical Interfaces}

\author{Jing Yang}
\email{j.yang@mpi-susmat.de}
\author{Bingxin Li}
\author{Samuel Mattoso}
\author{Ahmed Abdelkawy}
\author{Mira Todorova}
\email{m.todorova@mpi-susmat.de}
\author{J\"org Neugebauer}
\affiliation{Max Planck Institute for Sustainable Materials, Max-Planck-Str. 1, 40237, D\"usseldorf, Germany }

\begin{abstract}
Large thermal fluctuations of the liquid phase obscure the weak macroscopic electric field that drives electrochemical reactions, rendering the extraction of reliable interfacial charge distributions from \textit{ab initio} molecular dynamics extremely challenging. We introduce SMILE‑CP (Scalar Macro‑dipole Integrated LEarning – Charge Partitioning), a macro‑dipole‑constrained scheme that infers atomic charges using only the instantaneous atomic coordinates and the total dipole moment of the simulation cell --- quantities routinely available from standard density‑functional theory calculations. SMILE‑CP preserves both the global electrostatic field and the local potential without invoking any explicit charge‑partitioning scheme. Benchmarks on three representative electrochemical interfaces --- nanoconfined water, Mg$^{2+}$ dissolution in water, and a kinked Mg vicinal surface under anodic bias --- show that SMILE‑CP eliminates the qualitative errors observed for unconstrained charge decompositions. The method is computationally inexpensive and data‑efficient, opening the door to charge‑aware machine‑learning potentials capable of bias‑controlled, nanosecond‑scale simulations of realistic electrochemical systems.
\end{abstract}

\maketitle

Computational electrochemistry plays a crucial role in understanding and designing electrochemical processes at the atomic scale, including reactions at electrode interfaces, ion transport, and charge transfer phenomena \cite{magnussen2019toward, gonella2021water,ringe2021implicit,todorova2024first}. While density functional theory (DFT) provides an accurate quantum mechanical description, its high computational cost severely limits the accessible length and time scales, impeding 
simulations of realistic systems and dynamics. On the other hand, classical force fields enable larger-scale simulations but often rely on fixed or simplified electrostatic models, lacking the flexibility to capture complex, environment-dependent charge distributions that are needed for modeling electrochemical interfaces, especially charge transfer reactions.  

In recent years, machine learning interatomic potentials (MLIPs) have emerged as promising tools that combine near-DFT accuracy with significantly reduced computational cost \cite{deringer2019machine,zuo2020performance, unke2021machine}. However, the most prevalent types of MLIPs \cite{bartok2010gaussian, behler2021four, deringer2017machine, novikov2020mlip, drautz2019atomic, Bochkarev2022, yue2021short} determine the forces and energies of a given atom uniquely by its local atomistic environment within a certain cutoff. This inherent short-sightedness of the descriptors renders the models incapable of capturing long-range interactions accurately, limiting their applicability. Hence, incorporating long-range electrostatic interactions in MLIPs has been a major focus of ongoing research \cite{grisafi2019incorporating, unke2019physnet, ko2021fourth, zhang2022deep, Rinaldi2025, falletta2025unified,joll2024machine}. 

A key quantity in electrochemistry is the electrostatic potential \(\phi(z)\), which describes the variation of the electric potential as a function of position perpendicular to the electrode–electrolyte interface. This potential governs the distribution of ions, influences charge transfer processes, and plays a central role in determining interfacial properties such as the double layer structure and the driving force for electrochemical reactions. An accurate reproduction of $\phi$ is therefore essential for realistically modeling the behavior of electrochemical systems at the atomic scale. However, none of the present MLIP approaches takes the electrostatic potential $\phi$ directly as a target property due to its inherent incompatibility with standard ML framework. Unlike atomic energies and forces, $\phi$ is a continuous scalar field over 3D space computed as the solution to Poisson’s equation. Numerically, computing gradients of $\phi$ with respect to model parameters is impractical due to the nonlocality of the Poisson operator and therefore $\phi$ is ill-suited to be included in the loss function of a ML model. Instead, one commonly explored approach is to train the model on a local charge partition scheme and include the long-range electrostatic interactions computed from the local charge distribution. For example, Hirshfeld charges \cite{hirshfeld1977bonded} are used as reference for training the fourth-generation high-dimensional neutral network potentials \cite{ko2021fourth, shaidu2024incorporating, kocer2025iterative}. The deep potential (DP) model \cite{zhang2022deep} is trained on the centers of maximally localized Wannier functions (Wannier centers, WCs) \cite{marzari1997maximally,marzari2012maximally} to calculate the corresponding long-range interaction. These MLIPs are used for simulating systems containing charge species and electric field \cite{gao2022self, zhang2024molecular, zhu2024machine} and for modeling dielectric response in polarizable media \cite{Zhang2020, Krishnamoorthy2021, Han2023, liang2025polarizable}. 

In this work, we demonstrate that machine learning models that are solely trained to reproduce local charge decomposition schemes can result in significant errors in the long-range electrostatic potential $\phi(z)$. This shortcoming limits their application to electrochemical systems, in which charge transfer processes are strongly influenced by the macroscopic electric field. To address this issue, we introduce two innovations: (1) incorporating the macro dipole moment of the system as a surrogate for the electric field in the model's cost function to ensure accurate reconstruction of the long-range electrostatic profile and (2) explicitly accounting for the electronic polarization of water, which escapes standard machine learning potentials because of its relatively small magnitude in comparison to the intrinsic thermal fluctuation. The proposed approach is termed SMILE‑CP (Scalar Macro‑dipole Integrated LEarning – Charge Partitioning). 






\begin{figure}[th] 
\centering
\includegraphics[width=0.46\textwidth]{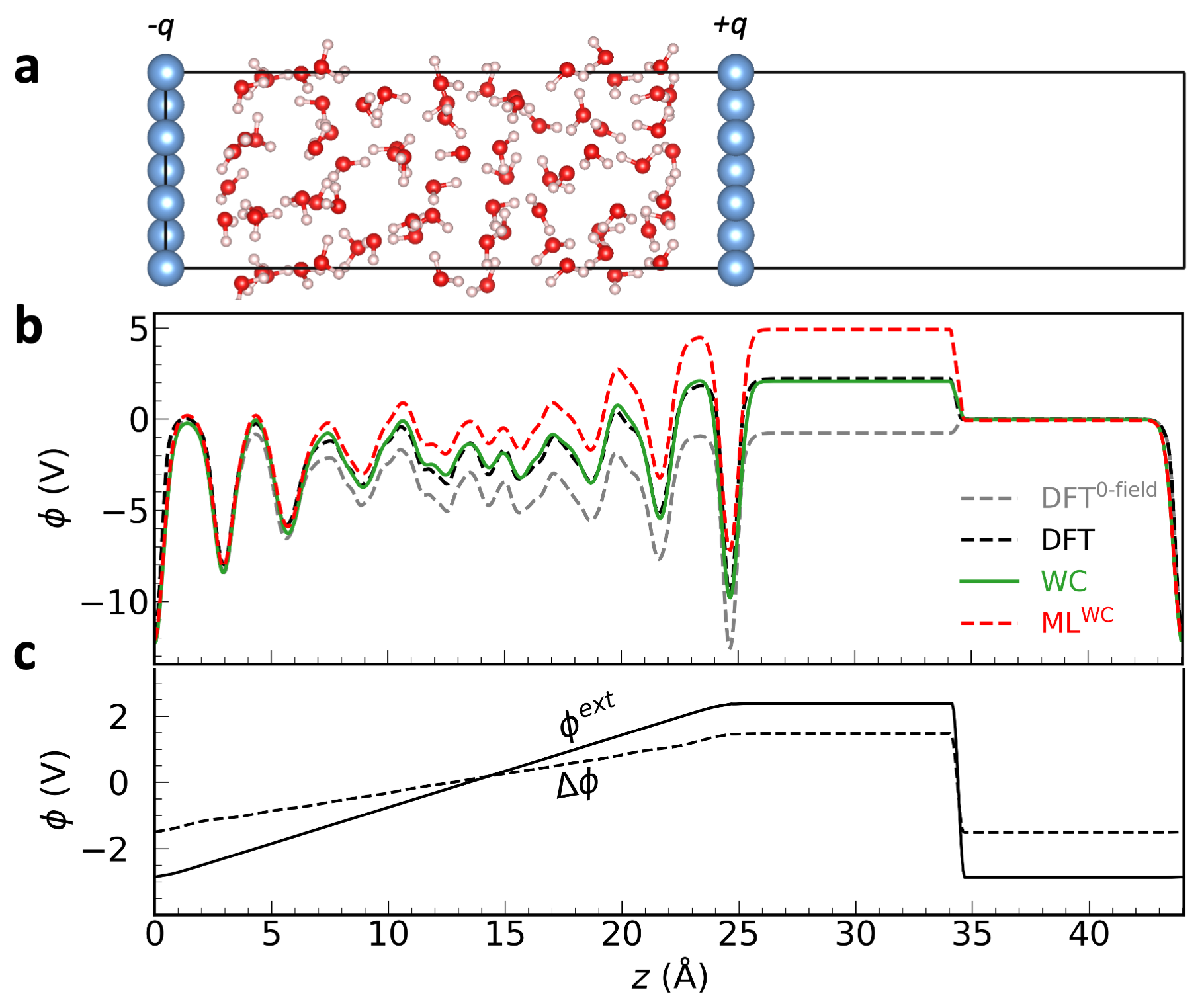}
\caption{\label{fig1} (a) Atomistic model of the nanoconfined water between two Ne electrodes. The two electrodes carry opposite charges, creating a macroscopic field across the water layer. (b) The corresponding electrostatic potential calculated from DFT, Wannier centers (WCs), and the ML model learning from individual WCs (ML$^{\mathrm{WC}}$), with an applied field of $E^\mathrm{ext} =$ 0.2 V/$\mathrm{\AA}$. The gray dashed line represents the zero field DFT reference. (c) The externally applied field $\phi^{\mathrm ext}$ and the difference between the DFT electrostatic potential with and without field $\Delta \phi$. In the bulk water region, $\Delta \phi$ corresponds to an electric field  smaller than $E^{\mathrm ext}$, which results from the electronic screening of water.     }
\end{figure}


To analyze and quantify the impact of large local fluctuations in the electrostatic potential on the accurate modeling of macroscopic fields in electrochemical systems, we have constructed a well-defined computational setup. This setup comprises a nanoconfined water slab that is sandwiched between two oppositely charged neon (Ne) electrodes (Fig. 1a)  \cite{Surendralal2018,Deissenbeck2021}. This setup realistically describes local fluctuations at an electrochemical interface while providing full quantitative access to and control over the macroscopic field. Thus, it provides an ideal testbed for evaluating the capacity of local ML models to capture macroscopic electrostatics amidst pronounced microscopic disorder, a challenge that is ubiquitous in realistic electrochemical interfaces. The net electrode charge, and thus the applied electric field across the water layer, is controlled by adjusting the Ne core charges while maintaining the cell's overall charge neutrality. We impose a  dipole correction along the surface normal ($z$) direction to ensure accurate long-range electrostatics \cite{Neugebauer1992}. \textit{Ab initio} molecular dynamics (AIMD) simulations are performed using the Vienna Ab Initio Simulation Package (VASP) \cite{vasp1,vasp2}. A charge decomposition is performed on a selected set of MD snapshots using Wannier90 \cite{mostofi2008wannier90}, and Bader charge decomposition \cite{bader2001properties, henkelman2006fast}; full computational details are provided in the supplemental material (SM) \cite{SM}.

Fig. \ref{fig1}b shows the electrostatic potential profile $\phi$, along the $z$ axis for a representative AIMD snapshot at an applied field of $E^\mathrm{ext} = $ 0.2 V/$\mathrm{\AA}$ (black line), compared with the zero-field case (grey dashed line). The close-to-linear dependence observed implies a nearly constant field in the water region. This is confirmed in Fig. \ref{fig1}c by plotting the difference potential, $\Delta\phi$, between the two field conditions. Bader charges fail to reproduce the macro dipole (see SM). Only the maximally localized Wannier functions accurately reproduce the macroscopic field from the WCs of individual water molecules (Fig. \ref{fig1}b, green line), while small local deviations reflect the Gaussian approximation of the electron density. 

In an ML approach, however, the WCs can no longer be obtained from electronic wave functions but exclusively from the local environment of each atom. We hence develop and apply a ML model (ML$^{\mathrm{WC}}$) based on local atomic cluster expansion (ACE) descriptors \cite{Drautz2019, lysogorskiy2021performant,Bochkarev2022,Lysogorskiy2023} which predicts the WC positions of individual water molecules, in the same spirit of the DP potential with long-range electrostatic interaction \cite{zhang2022deep}. The ML$^{\mathrm{WC}}$ model achieves state-of-the-art accuracy in WC positions with a mean squared root error (RMSE) of 0.004 $\mathrm{\AA}$ \cite{zhang2024molecular,zhu2024machine}. However, despite this impressive local accuracy of 4 pm, the model systematically fails to reproduce the correct macroscopic field response, instead predicting an offset in the potential (Fig. \ref{fig1}b, red line). This failure is generic: it arises because large local fluctuations, which are inherent to electrochemical systems, decouple precise local charge prediction from accurate long-range electrostatics. Our finding thus reveals a fundamental limitation of strictly local ML approaches in capturing macroscopic electrostatic behavior in complex interfacial systems. 

\begin{figure}[h] 
\centering
\includegraphics[width=0.45\textwidth]{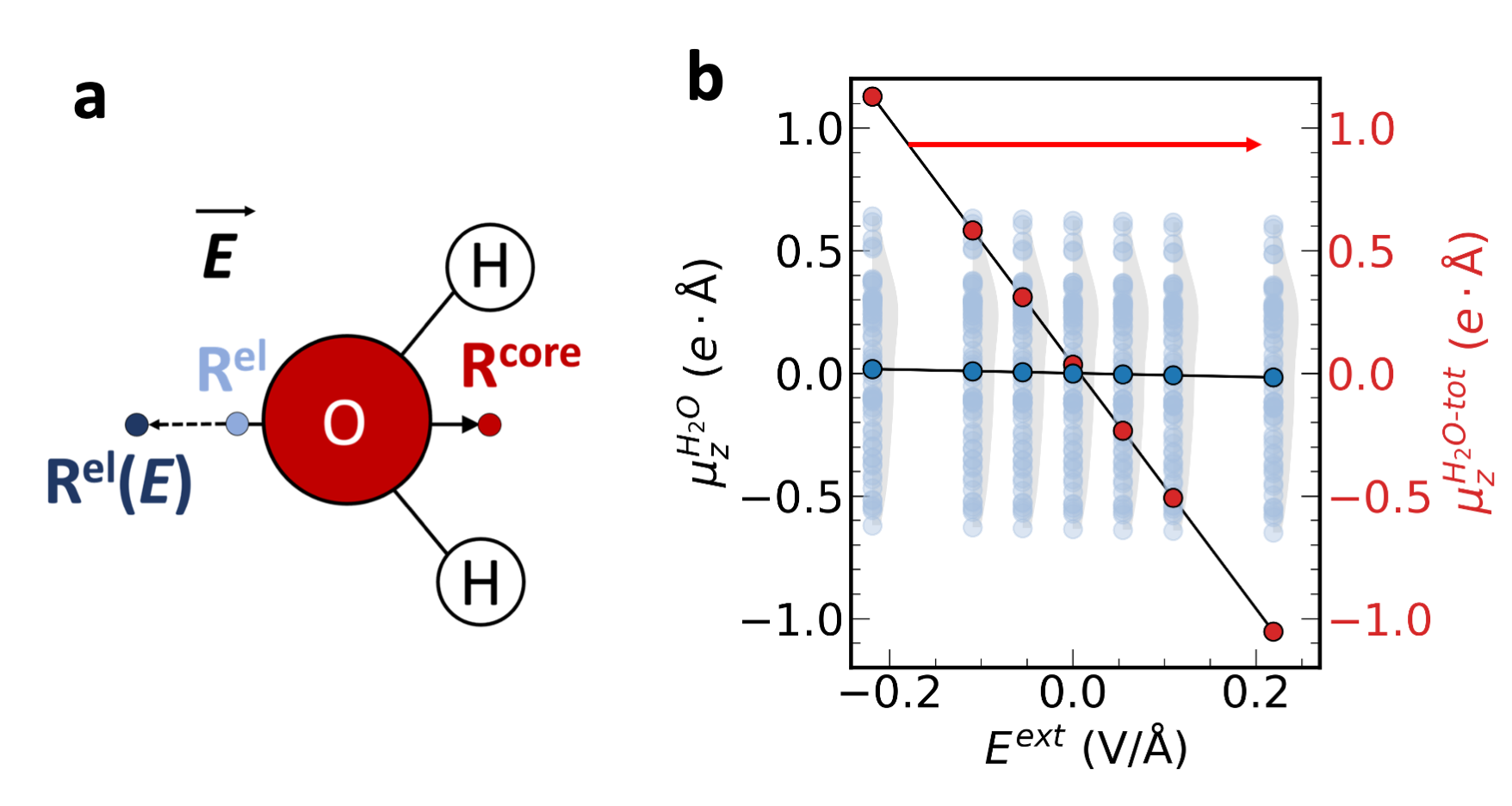}
\caption{\label{fig2} (a) Schematic showing how a water molecule responds to an external electric field. The dipole moment of individual water molecule is represented by $\boldsymbol{\mu}^\mathrm{H_2O} = 8e(\mathbf{R^{\mathrm{el}}} - \mathbf{R^{\mathrm{core}}})$, where $\mathbf{R^{\mathrm{core}}}$ is the core charge center and $\mathbf{R^{\mathrm{el}}}$ is the electron charge center. When an external field is applied, the electrons are polarized, leading to change of $\mathbf{R^{\mathrm{el}}}$ and thus $\boldsymbol{\mu}^\mathrm{H_2O}$.  (b) The dipole moments in the $z$ direction of individual water molecules $\mu_z^\mathrm{H_2O}$ (light blue dots) as a function of the applied field strength $E^\mathrm{ext}$. The corresponding grey shades show their value distribution. The red dots are the sum of the dipole moment in the $z$ direction $\mu^\mathrm{H_2O\text{-}tot}_z$, which clearly shows a linear relationship with $E^\mathrm{ext}$. The dark blue dots show $\mu^\mathrm{H_2O\text{-}tot}_z$ averaged to each water molecule, which is almost horizontal on the given scale.     }
\end{figure}

To elucidate the origin of this failure, we analyze how individual water dipoles respond to the applied electric field. Specifically, we define the electron charge center of each water molecule as the center of mass of its WCs ($\mathbf{R}^\mathrm{el}$), and calculate the molecular dipole moment as $\boldsymbol{\mu}^\mathrm{H_2O} = 8e(\mathbf{R^{\mathrm{el}}} - \mathbf{R^{\mathrm{core}}})$, where $\mathbf{R^{\mathrm{core}}}$ denotes the core position (see schematic in Fig. \ref{fig2}a). Upon application of an external field, the electron cloud of each water molecule is polarized, leading to a shift in the WC positions and a corresponding change in  $\boldsymbol{\mu}^\mathrm{H_2O}$. 

By tracking changes in the WC positions under varying external fields, we compute the dipole moment distributions for all water molecules. Fig. \ref{fig2}b displays both the individual dipole moment projections along $z$ ($\mu_z^\mathrm{H_2O}$, light blue dots) and their sum ($\mu_z^\mathrm{H_2O\text{-}tot}$, red dots) as a function of the applied field $E^\mathrm{ext}$. While the total dipole response exhibits a clear linear dependence on $E^\mathrm{ext}$---in agreement with the expected global polarization---the distributions of individual $\mu_z^\mathrm{H_2O}$ remain essentially unchanged. This is further underscored by the nearly horizontal trend of the per-molecule-averaged $\mu_z^\mathrm{H_2O}$ (dark blue dots in Fig. \ref{fig2}b).

This seemingly contradictory observation can be explained quantitatively. For a net change of 1\,$e\cdot\mathrm{\AA}$ in the global dipole moment distributed across the 64 water molecules in our system, the average change in the individual molecular dipole $\mu_z^\mathrm{H_2O}$ is just 0.016\,$e\cdot\mathrm{\AA}$, two orders of magnitude smaller than the thermal fluctuations in $\mu_z^\mathrm{H_2O}$present in the system. This disparity presents a fundamental challenge for ML models trained to predict $\mathbf{R}^\mathrm{el}$: the relevant signal---the material's response to an external field---is almost entirely obscured by the intrinsic noise due to thermal motion and orientational disorder of water molecules. For WCs, this translates to a change of only 0.002 $\mathrm{\AA}$ change in $\mathbf{R}^\mathrm{el}$, well below the resolution achievable by local ML models. This limitation persists even at zero external field, due to the long-range field arising from spontaneous dipole fluctuations, and cannot be cured by adjusting model hyperparameters (see SM \cite{SM}).

Since the local descriptor entering the ML$^{\mathrm{WC}}$ model is not explicitly dependent on the macroscopic field, these models can only describe the local ionic screening but fail to capture the electronic polarization. Thus, in the absence of ionic screening (i.e. freezing in an AIMD snapshot) and applying an external field, the effective field in the water region $E_{\mathrm{water}}\approx E^\mathrm{ext}$. Indeed, as shown in Fig. \ref{fig1}b, the internal electric field predicted by ML$^{\mathrm{WC}}$ is nearly twice that from DFT, corresponding to the missing dielectric screening factor ($\varepsilon_{\infty} \approx 2$). These limitations apply generally to all ML models that use local charges as input. 
  
\begin{figure}[h] 
\centering
\includegraphics[width=0.45\textwidth]{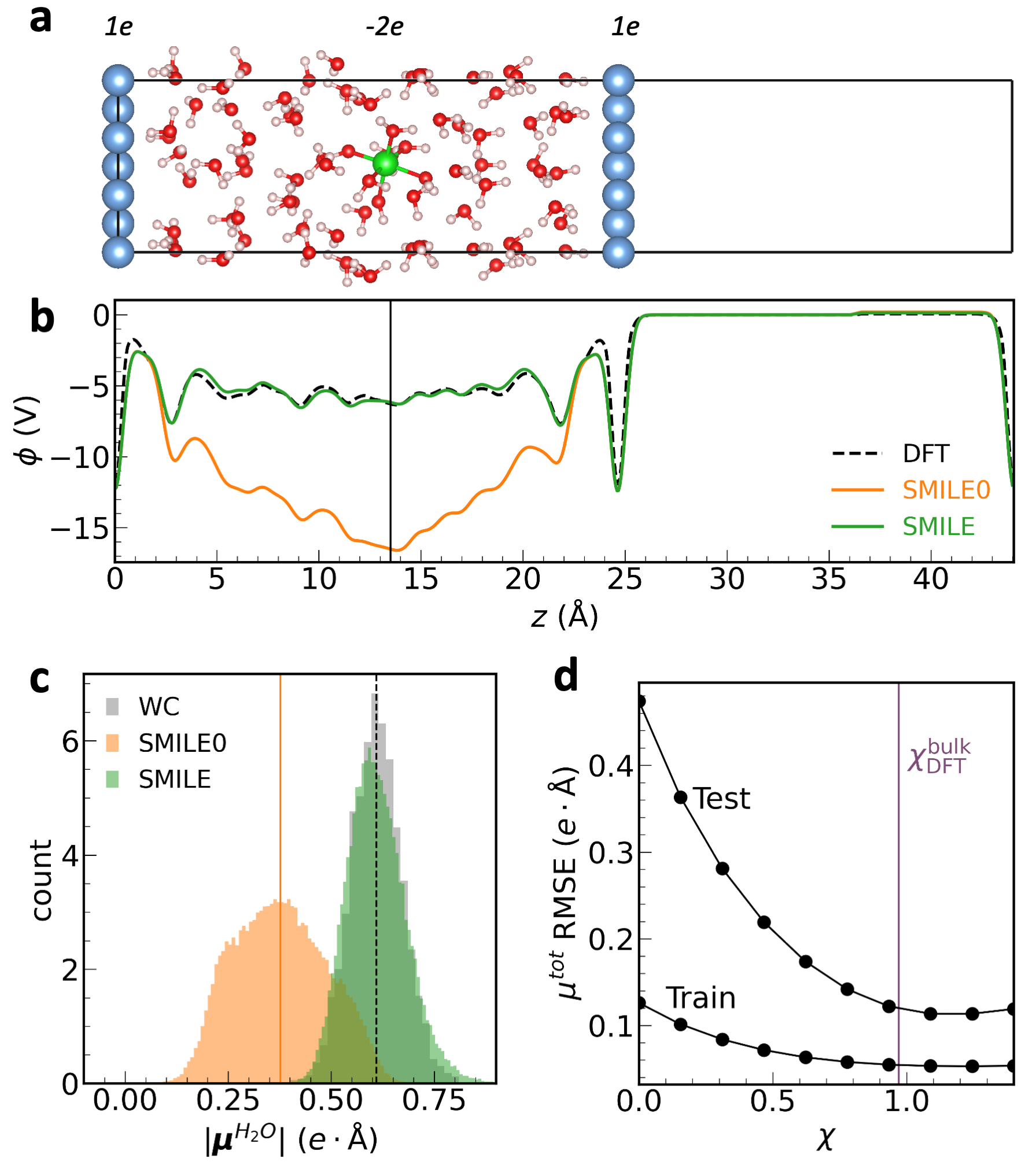}
\caption{\label{fig3} (a) Atomistic structure of nanoconfined water with a dissolved Mg$^{2+}$ ion. The compensating charge of the 2+ ion is evenly spread on the two Ne electrodes. (b) Predicted electrostatic potential $\phi$ of the SMILE0 and SMILE model, in comparison with DFT. The SMILE0 model is equivalent to SMILE at $\chi$ = 0. The black vertical line represents the average position of the Mg ion. (c) The distribution of the individual water molecule dipole magnitude $|\boldsymbol{\mu}^\mathrm{H_2O}|$ of the two models, in comparison to the values computed with WCs. The vertical lines show the corresponding averaged values, for which the lines of WC and SMILE overlap. (d) The RMSE of the SMILE model as a function of $\chi$. The purple vertical line shows the DFT-calculated electronic susceptibility of water $\chi_{\mathrm{DFT}}^{\mathrm{bulk}}$ = 0.96.}
\end{figure}

Given the failure of the local ML model to capture the long-range electrostatics, we propose an approach that extracts the local charges directly from the total macroscopic dipole moment $\mu^\mathrm{tot}_z$. In this approach, the dipole moment of each water molecule $\boldsymbol{\mu_i}$ is a function of its local descriptor $\mathbf{D}^i$, such that 
\begin{equation}
    \boldsymbol{\mu}^{i} = f(\mathbf{D}^i).
    \label{eq:model}
\end{equation} 
The sum of the predicted molecular dipoles has to match the total dipole along $z$ axis:
\begin{equation}
    \sum_i\mu^i_z = \mu_z^\mathrm{tot},
    \label{eq:macro}
\end{equation}
analogous to the procedure used in MLIPs, where atomic energies are learned subject to the constraint $\sum_i E^i = E^\mathrm{tot}$. We denote the ML model employing Eq. \ref{eq:macro} as SMILE0. Since in this approach the charge decomposition is done with the condition to minimize errors in the macro dipole, this schema guarantees an accurate description of the macroscopic field. 

However, when applying this approach to a more realistic system with dissolved Mg$^{2+}$ ion and charged electrodes (Fig. \ref{fig3}a), severe deviations between the electrostatic potential obtained by DFT (Fig. \ref{fig3}b, black line) and the SMILE0 model (Fig. \ref{fig3}b, orange line) are observed: while DFT shows strong screening as a nearly flat potential in the long range, the SMILE0 model predicts a pronounced field and a valley-shaped potential, indicative of underestimated screening.

This failure arises from the absence of electronic polarization in typical ML models. Conceptually, the total macro dipole can be separated into an electronic polarization term $\mu_z^\mathrm{el\text{-}pol}$, that is field-induced, and a local term $\mu_i^\mathrm{local}$, that is field-independent:
\begin{equation}
\mu^\mathrm{tot}_z  = \mu^\mathrm{el\text{-}pol}_z + \mu_z^\mathrm{local}.
\label{eq:sum}
\end{equation}
In the linear response regime, $\mu_z^\mathrm{el\text{-}pol} = -\chi \mu_z^\mathrm{tot}$, with $\chi$ being the electronic susceptibility ($\chi = \varepsilon_\infty-1$). The induced electronic polarization dipole counterbalances the applied field, therefore $\mu_z^\mathrm{el\text{-}pol}$ and $\mu_z^\mathrm{tot}$ have opposite signs. Thus, $\mu^\mathrm{tot}_z = \mu^\mathrm{local}_z/(1+\chi)$, showing that the observed macro dipole is only a screened fraction of the static local dipoles, leading to a systematic underestimation of screening. Fig. \ref{fig3}c confirms this, with the DFT-derived WC dipole distribution averaged at 0.61 $e\cdot \mathrm{\AA}$, while SMILE0 yields 0.38 $e\cdot \mathrm{\AA}$. 

To account for the electronic polarization, we exploit the fact that in contrast to the ionic dielectric constant, which shows huge spatial fluctuations at the electrochemical interface, electronic polarization is remarkably homogeneous (see Fig. \ref{fig1}c). Based on this insight, we reformulate the constraint in Eq. \ref{eq:macro} as
\begin{equation}
   \sum_i \mu^{i, \mathrm{local}}_z = (1+\chi)\mu^\mathrm{tot}_z,
   \label{eq:local}
\end{equation}
 and denote this model as SMILE. Varying $\chi$, we find the RMSE between reference and predicted dipoles is minimized at $\chi$ $\approx$ 1.2 (Fig. \ref{fig3}d), consistent across several water-solid interface systems. This optimal value slightly exceeds the DFT-calculated bulk value ($\chi_{\mathrm{DFT}}^{\mathrm{bulk}}$ = 0.96), likely due to a reduced HOMO-LUMO gap of water in the interface region resulting in enhanced dielectric response \cite{PhD_Surendralal}.

 With this correction, the SMILE model faithfully reproduces both the local dipole distribution (average 0.61 $e\cdot \mathrm{\AA}$, Fig. \ref{fig3}c) and the long-range electrostatics (Fig. \ref{fig3}b, green line), achieving an RMSE of 0.046 $e\cdot \mathrm{\AA}$ for the macro dipole and 0.056 $e\cdot \mathrm{\AA}$ for individual water dipoles. These results demonstrate the model's strong predictive power for both local and macroscopic electrostatics, resolving the inherent shortcomings of strictly local ML approaches. 
 
\begin{figure}[h] 
\centering
\includegraphics[width=0.45\textwidth]{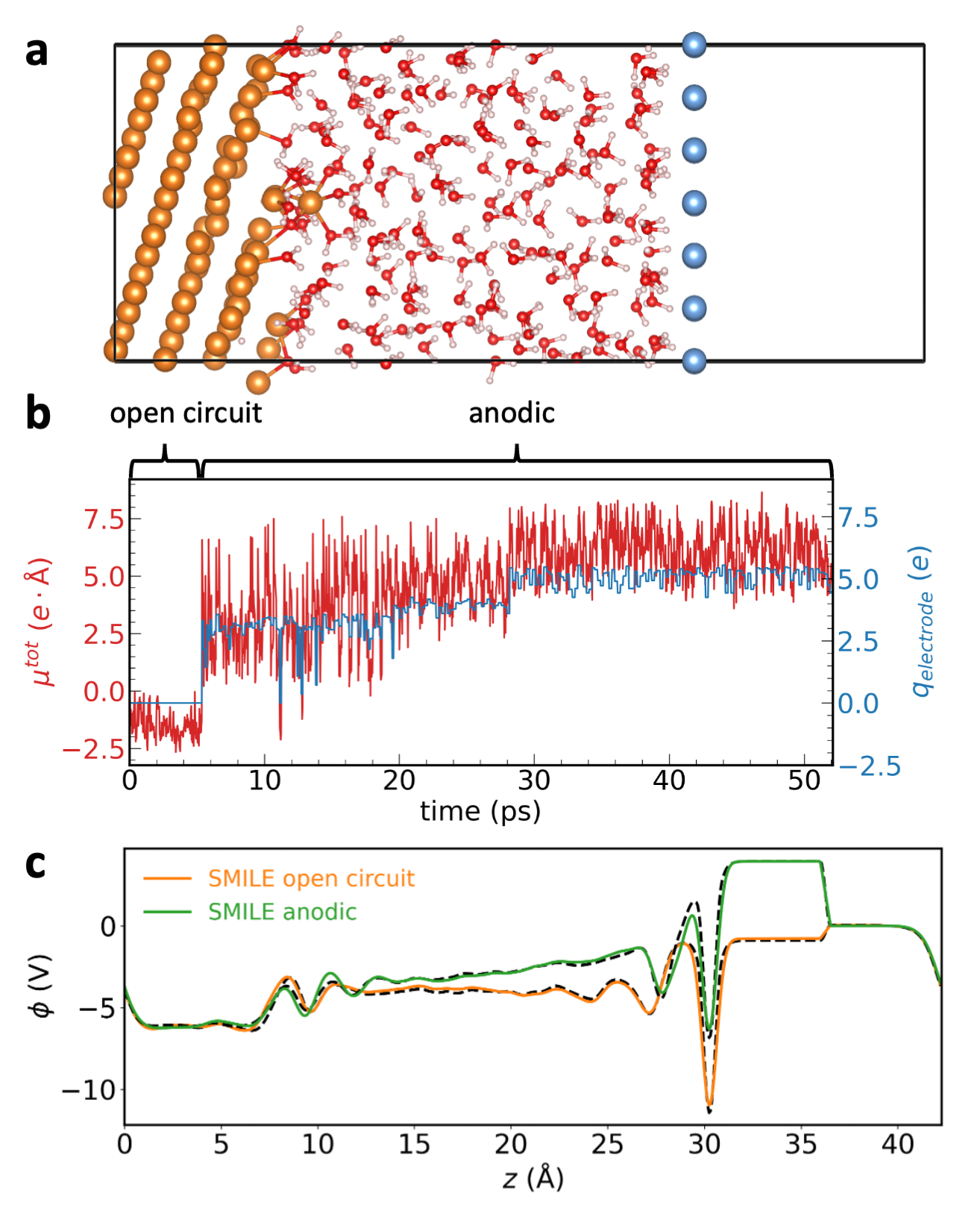}
\caption{\label{fig4} (a) Atomistic model of the Mg(12$\bar{3}$5)/water interface. (b) The evolution of the macro dipole $\mu^\mathrm{tot}$ and the electrode charge $q_{\mathrm{electrode}}$ as a function of time in the AIMD simulation. The first 5 ps is under open circuit condition, followed by 45 ps of anodic bias. (c) The predicted electrostatic potential $\phi$ under open circuit condition and anodic bias predicted by the SMILE model (colored lines) in comparison to the DFT references (black dashed lines).  }
\end{figure}

Another advantage of the proposed model is that it is flexible for learning the local dipole or the local charge. The latter assigns a point charge to each atom in the system. Practically, this is done by rewriting Eq. \ref{eq:model} and \ref{eq:local} as
\begin{equation}
    q_i = f(\mathbf{D}^i),
\end{equation}
\begin{equation}
   \sum_i q_iz_i = (1+\chi)\mu^\mathrm{tot}_z. 
   \label{eq:q}
\end{equation}
Here $q_i$ is the charge on atom $i$ and $z_i$ is the $z$ coordinate of the corresponding atom. 
The model assigning atomic charges is arguably more flexible than the model predicting molecular dipoles, because by assigning a dipole moment to each water molecule, the dipole-based model implicitly assumes charge neutrality for the individual molecule, thereby excluding intermolecular charge transfer. 

We apply the model employing Eq. \ref{eq:q} to a realistic electrochemical system, a Mg(12$\bar{3}$5) vicinal surface under anodic potential (Fig. \ref{fig4}a). The Mg/water interface is of particular interest due to the observed anomalous anodic hydrogen evolution reaction \cite{beetz1866xxxiv,song1999corrosion}, and in recent years there have been a number of computational studies attempting to reveal the reaction pathway \cite{Surendralal2018, yuwono2019understanding, deissenbeck2024revealing, li2024uncovering}. In this study, we employ a static potentiostat to apply anodic bias on the Mg surface, similar to the method in \cite{Surendralal2018}. The evolution of the macro dipole and the electrode charge is shown in Fig. \ref{fig4}b. The first 5 ps of the AIMD run is under open circuit condition, with no charge on the electrode, then an anodic bias is applied, which is gradually ramped up to 4 V.

The SMILE model consistently reproduces the electrostatic potential both under open circuit condition and under anodic bias (Fig. \ref{fig4}c). The long AIMD trajectory allows us to average out the local fluctuations in $\phi$ and clearly observe the macroscopic field across the cell. While the $\phi$ profile is almost flat in the bulk of water under open circuit condition, the field under the 4V anodic bias averages to 0.15 V/$\mathrm{\AA}$. These long-range macroscopic fields are accurately captured by the SMILE model. We note here that both curves are obtained with a single model, which highlights the transferability of the model across applied potentials.    
 
In conclusion, for typical electrochemical interfaces, the polarization induced by an applied electric field is roughly two orders of magnitude smaller than the thermal dipole fluctuations of the liquid. As a result, machine‑learning potentials that are trained only on locally partitioned charges or dipoles tend to fit the noisy local variations and effectively discard the much weaker macroscopic component, which leads to unacceptably large errors in the long‑range electrostatic field. Our macro‑dipole‑constrained SMILE-CP approach restores the correct field while preserving the accuracy of the local charge distribution. The only additional information required is the scalar total dipole, which is readily obtainable from standard AIMD runs. Thus, the method avoids any ambiguous charge‑partitioning step. Because the constraint is introduced as a loss‑function term, it works with any descriptor and any neural‑network architecture. Benchmarks on nanoconfined water, Mg$^{2+}$ dissolution, and a biased Mg vicinal surface show that the macro‑dipole constraint eliminates the qualitative errors in the electrostatic potential observed for unconstrained MLIPs and reproduces the full Poisson solution from the scalar charges alone. This flexible, easy‑to‑implement, and data‑efficient strategy therefore opens the door to bias‑controlled, nanosecond‑scale simulations of realistic electrochemical interfaces and paves the way for systematic investigations of voltage‑dependent processes in batteries, fuel cells, and electrocatalysis.


We acknowledge funding by the Deutsche Forschungsgemeinschaft (DFG, German Research Foundation) through SFB1394, project no. 409476157 and SFB1625, project no. 506711657. J. Y. acknowledge support by the Alexander von Humboldt foundation.

\bibliography{reference}

\begin{thebibliography}{51}%
\makeatletter
\providecommand \@ifxundefined [1]{%
 \@ifx{#1\undefined}
}%
\providecommand \@ifnum [1]{%
 \ifnum #1\expandafter \@firstoftwo
 \else \expandafter \@secondoftwo
 \fi
}%
\providecommand \@ifx [1]{%
 \ifx #1\expandafter \@firstoftwo
 \else \expandafter \@secondoftwo
 \fi
}%
\providecommand \natexlab [1]{#1}%
\providecommand \enquote  [1]{``#1''}%
\providecommand \bibnamefont  [1]{#1}%
\providecommand \bibfnamefont [1]{#1}%
\providecommand \citenamefont [1]{#1}%
\providecommand \href@noop [0]{\@secondoftwo}%
\providecommand \href [0]{\begingroup \@sanitize@url \@href}%
\providecommand \@href[1]{\@@startlink{#1}\@@href}%
\providecommand \@@href[1]{\endgroup#1\@@endlink}%
\providecommand \@sanitize@url [0]{\catcode `\\12\catcode `\$12\catcode
  `\&12\catcode `\#12\catcode `\^12\catcode `\_12\catcode `\%12\relax}%
\providecommand \@@startlink[1]{}%
\providecommand \@@endlink[0]{}%
\providecommand \url  [0]{\begingroup\@sanitize@url \@url }%
\providecommand \@url [1]{\endgroup\@href {#1}{\urlprefix }}%
\providecommand \urlprefix  [0]{URL }%
\providecommand \Eprint [0]{\href }%
\providecommand \doibase [0]{https://doi.org/}%
\providecommand \selectlanguage [0]{\@gobble}%
\providecommand \bibinfo  [0]{\@secondoftwo}%
\providecommand \bibfield  [0]{\@secondoftwo}%
\providecommand \translation [1]{[#1]}%
\providecommand \BibitemOpen [0]{}%
\providecommand \bibitemStop [0]{}%
\providecommand \bibitemNoStop [0]{.\EOS\space}%
\providecommand \EOS [0]{\spacefactor3000\relax}%
\providecommand \BibitemShut  [1]{\csname bibitem#1\endcsname}%
\let\auto@bib@innerbib\@empty
\bibitem [{\citenamefont {Magnussen}\ and\ \citenamefont
  {Gro{\ss}}(2019)}]{magnussen2019toward}%
  \BibitemOpen
  \bibfield  {author} {\bibinfo {author} {\bibfnamefont {O.~M.}\ \bibnamefont
  {Magnussen}}\ and\ \bibinfo {author} {\bibfnamefont {A.}~\bibnamefont
  {Gro{\ss}}},\ }\bibfield  {title} {\bibinfo {title} {Toward an atomic-scale
  understanding of electrochemical interface structure and dynamics},\
  }\href@noop {} {\bibfield  {journal} {\bibinfo  {journal} {Journal of the
  American Chemical Society}\ }\textbf {\bibinfo {volume} {141}},\ \bibinfo
  {pages} {4777} (\bibinfo {year} {2019})}\BibitemShut {NoStop}%
\bibitem [{\citenamefont {Gonella}\ \emph {et~al.}(2021)\citenamefont
  {Gonella}, \citenamefont {Backus}, \citenamefont {Nagata}, \citenamefont
  {Bonthuis}, \citenamefont {Loche}, \citenamefont {Schlaich}, \citenamefont
  {Netz}, \citenamefont {K{\"u}hnle}, \citenamefont {McCrum}, \citenamefont
  {Koper} \emph {et~al.}}]{gonella2021water}%
  \BibitemOpen
  \bibfield  {author} {\bibinfo {author} {\bibfnamefont {G.}~\bibnamefont
  {Gonella}}, \bibinfo {author} {\bibfnamefont {E.~H.}\ \bibnamefont {Backus}},
  \bibinfo {author} {\bibfnamefont {Y.}~\bibnamefont {Nagata}}, \bibinfo
  {author} {\bibfnamefont {D.~J.}\ \bibnamefont {Bonthuis}}, \bibinfo {author}
  {\bibfnamefont {P.}~\bibnamefont {Loche}}, \bibinfo {author} {\bibfnamefont
  {A.}~\bibnamefont {Schlaich}}, \bibinfo {author} {\bibfnamefont {R.~R.}\
  \bibnamefont {Netz}}, \bibinfo {author} {\bibfnamefont {A.}~\bibnamefont
  {K{\"u}hnle}}, \bibinfo {author} {\bibfnamefont {I.~T.}\ \bibnamefont
  {McCrum}}, \bibinfo {author} {\bibfnamefont {M.~T.}\ \bibnamefont {Koper}},
  \emph {et~al.},\ }\bibfield  {title} {\bibinfo {title} {Water at charged
  interfaces},\ }\href@noop {} {\bibfield  {journal} {\bibinfo  {journal}
  {Nature Reviews Chemistry}\ }\textbf {\bibinfo {volume} {5}},\ \bibinfo
  {pages} {466} (\bibinfo {year} {2021})}\BibitemShut {NoStop}%
\bibitem [{\citenamefont {Ringe}\ \emph {et~al.}(2021)\citenamefont {Ringe},
  \citenamefont {Hormann}, \citenamefont {Oberhofer},\ and\ \citenamefont
  {Reuter}}]{ringe2021implicit}%
  \BibitemOpen
  \bibfield  {author} {\bibinfo {author} {\bibfnamefont {S.}~\bibnamefont
  {Ringe}}, \bibinfo {author} {\bibfnamefont {N.~G.}\ \bibnamefont {Hormann}},
  \bibinfo {author} {\bibfnamefont {H.}~\bibnamefont {Oberhofer}},\ and\
  \bibinfo {author} {\bibfnamefont {K.}~\bibnamefont {Reuter}},\ }\bibfield
  {title} {\bibinfo {title} {Implicit solvation methods for catalysis at
  electrified interfaces},\ }\href@noop {} {\bibfield  {journal} {\bibinfo
  {journal} {Chemical Reviews}\ }\textbf {\bibinfo {volume} {122}},\ \bibinfo
  {pages} {10777} (\bibinfo {year} {2021})}\BibitemShut {NoStop}%
\bibitem [{\citenamefont {Todorova}\ \emph {et~al.}(2024)\citenamefont
  {Todorova}, \citenamefont {Wippermann},\ and\ \citenamefont
  {Neugebauer}}]{todorova2024first}%
  \BibitemOpen
  \bibfield  {author} {\bibinfo {author} {\bibfnamefont {M.}~\bibnamefont
  {Todorova}}, \bibinfo {author} {\bibfnamefont {S.}~\bibnamefont
  {Wippermann}},\ and\ \bibinfo {author} {\bibfnamefont {J.}~\bibnamefont
  {Neugebauer}},\ }\bibfield  {title} {\bibinfo {title} {First principles
  approaches and concepts for electrochemical systems},\ }\href@noop {}
  {\bibfield  {journal} {\bibinfo  {journal} {arXiv preprint arXiv:2411.05925}\
  } (\bibinfo {year} {2024})}\BibitemShut {NoStop}%
\bibitem [{\citenamefont {Deringer}\ \emph {et~al.}(2019)\citenamefont
  {Deringer}, \citenamefont {Caro},\ and\ \citenamefont
  {Cs{\'a}nyi}}]{deringer2019machine}%
  \BibitemOpen
  \bibfield  {author} {\bibinfo {author} {\bibfnamefont {V.~L.}\ \bibnamefont
  {Deringer}}, \bibinfo {author} {\bibfnamefont {M.~A.}\ \bibnamefont {Caro}},\
  and\ \bibinfo {author} {\bibfnamefont {G.}~\bibnamefont {Cs{\'a}nyi}},\
  }\bibfield  {title} {\bibinfo {title} {Machine learning interatomic
  potentials as emerging tools for materials science},\ }\href@noop {}
  {\bibfield  {journal} {\bibinfo  {journal} {Advanced Materials}\ }\textbf
  {\bibinfo {volume} {31}},\ \bibinfo {pages} {1902765} (\bibinfo {year}
  {2019})}\BibitemShut {NoStop}%
\bibitem [{\citenamefont {Zuo}\ \emph {et~al.}(2020)\citenamefont {Zuo},
  \citenamefont {Chen}, \citenamefont {Li}, \citenamefont {Deng}, \citenamefont
  {Chen}, \citenamefont {Behler}, \citenamefont {Cs{\'a}nyi}, \citenamefont
  {Shapeev}, \citenamefont {Thompson}, \citenamefont {Wood} \emph
  {et~al.}}]{zuo2020performance}%
  \BibitemOpen
  \bibfield  {author} {\bibinfo {author} {\bibfnamefont {Y.}~\bibnamefont
  {Zuo}}, \bibinfo {author} {\bibfnamefont {C.}~\bibnamefont {Chen}}, \bibinfo
  {author} {\bibfnamefont {X.}~\bibnamefont {Li}}, \bibinfo {author}
  {\bibfnamefont {Z.}~\bibnamefont {Deng}}, \bibinfo {author} {\bibfnamefont
  {Y.}~\bibnamefont {Chen}}, \bibinfo {author} {\bibfnamefont {J.}~\bibnamefont
  {Behler}}, \bibinfo {author} {\bibfnamefont {G.}~\bibnamefont {Cs{\'a}nyi}},
  \bibinfo {author} {\bibfnamefont {A.~V.}\ \bibnamefont {Shapeev}}, \bibinfo
  {author} {\bibfnamefont {A.~P.}\ \bibnamefont {Thompson}}, \bibinfo {author}
  {\bibfnamefont {M.~A.}\ \bibnamefont {Wood}}, \emph {et~al.},\ }\bibfield
  {title} {\bibinfo {title} {Performance and cost assessment of machine
  learning interatomic potentials},\ }\href@noop {} {\bibfield  {journal}
  {\bibinfo  {journal} {The Journal of Physical Chemistry A}\ }\textbf
  {\bibinfo {volume} {124}},\ \bibinfo {pages} {731} (\bibinfo {year}
  {2020})}\BibitemShut {NoStop}%
\bibitem [{\citenamefont {Unke}\ \emph {et~al.}(2021)\citenamefont {Unke},
  \citenamefont {Chmiela}, \citenamefont {Sauceda}, \citenamefont {Gastegger},
  \citenamefont {Poltavsky}, \citenamefont {Schutt}, \citenamefont
  {Tkatchenko},\ and\ \citenamefont {Muller}}]{unke2021machine}%
  \BibitemOpen
  \bibfield  {author} {\bibinfo {author} {\bibfnamefont {O.~T.}\ \bibnamefont
  {Unke}}, \bibinfo {author} {\bibfnamefont {S.}~\bibnamefont {Chmiela}},
  \bibinfo {author} {\bibfnamefont {H.~E.}\ \bibnamefont {Sauceda}}, \bibinfo
  {author} {\bibfnamefont {M.}~\bibnamefont {Gastegger}}, \bibinfo {author}
  {\bibfnamefont {I.}~\bibnamefont {Poltavsky}}, \bibinfo {author}
  {\bibfnamefont {K.~T.}\ \bibnamefont {Schutt}}, \bibinfo {author}
  {\bibfnamefont {A.}~\bibnamefont {Tkatchenko}},\ and\ \bibinfo {author}
  {\bibfnamefont {K.-R.}\ \bibnamefont {Muller}},\ }\bibfield  {title}
  {\bibinfo {title} {Machine learning force fields},\ }\href@noop {} {\bibfield
   {journal} {\bibinfo  {journal} {Chemical Reviews}\ }\textbf {\bibinfo
  {volume} {121}},\ \bibinfo {pages} {10142} (\bibinfo {year}
  {2021})}\BibitemShut {NoStop}%
\bibitem [{\citenamefont {Bart{\'o}k}\ \emph {et~al.}(2010)\citenamefont
  {Bart{\'o}k}, \citenamefont {Payne}, \citenamefont {Kondor},\ and\
  \citenamefont {Cs{\'a}nyi}}]{bartok2010gaussian}%
  \BibitemOpen
  \bibfield  {author} {\bibinfo {author} {\bibfnamefont {A.~P.}\ \bibnamefont
  {Bart{\'o}k}}, \bibinfo {author} {\bibfnamefont {M.~C.}\ \bibnamefont
  {Payne}}, \bibinfo {author} {\bibfnamefont {R.}~\bibnamefont {Kondor}},\ and\
  \bibinfo {author} {\bibfnamefont {G.}~\bibnamefont {Cs{\'a}nyi}},\ }\bibfield
   {title} {\bibinfo {title} {Gaussian approximation potentials: The accuracy
  of quantum mechanics, without the electrons},\ }\href@noop {} {\bibfield
  {journal} {\bibinfo  {journal} {Physical review letters}\ }\textbf {\bibinfo
  {volume} {104}},\ \bibinfo {pages} {136403} (\bibinfo {year}
  {2010})}\BibitemShut {NoStop}%
\bibitem [{\citenamefont {Behler}(2021)}]{behler2021four}%
  \BibitemOpen
  \bibfield  {author} {\bibinfo {author} {\bibfnamefont {J.}~\bibnamefont
  {Behler}},\ }\bibfield  {title} {\bibinfo {title} {Four generations of
  high-dimensional neural network potentials},\ }\href@noop {} {\bibfield
  {journal} {\bibinfo  {journal} {Chemical Reviews}\ }\textbf {\bibinfo
  {volume} {121}},\ \bibinfo {pages} {10037} (\bibinfo {year}
  {2021})}\BibitemShut {NoStop}%
\bibitem [{\citenamefont {Deringer}\ and\ \citenamefont
  {Cs{\'a}nyi}(2017)}]{deringer2017machine}%
  \BibitemOpen
  \bibfield  {author} {\bibinfo {author} {\bibfnamefont {V.~L.}\ \bibnamefont
  {Deringer}}\ and\ \bibinfo {author} {\bibfnamefont {G.}~\bibnamefont
  {Cs{\'a}nyi}},\ }\bibfield  {title} {\bibinfo {title} {Machine learning based
  interatomic potential for amorphous carbon},\ }\href@noop {} {\bibfield
  {journal} {\bibinfo  {journal} {Physical Review B}\ }\textbf {\bibinfo
  {volume} {95}},\ \bibinfo {pages} {094203} (\bibinfo {year}
  {2017})}\BibitemShut {NoStop}%
\bibitem [{\citenamefont {Novikov}\ \emph {et~al.}(2020)\citenamefont
  {Novikov}, \citenamefont {Gubaev}, \citenamefont {Podryabinkin},\ and\
  \citenamefont {Shapeev}}]{novikov2020mlip}%
  \BibitemOpen
  \bibfield  {author} {\bibinfo {author} {\bibfnamefont {I.~S.}\ \bibnamefont
  {Novikov}}, \bibinfo {author} {\bibfnamefont {K.}~\bibnamefont {Gubaev}},
  \bibinfo {author} {\bibfnamefont {E.~V.}\ \bibnamefont {Podryabinkin}},\ and\
  \bibinfo {author} {\bibfnamefont {A.~V.}\ \bibnamefont {Shapeev}},\
  }\bibfield  {title} {\bibinfo {title} {The mlip package: moment tensor
  potentials with mpi and active learning},\ }\href@noop {} {\bibfield
  {journal} {\bibinfo  {journal} {Machine Learning: Science and Technology}\
  }\textbf {\bibinfo {volume} {2}},\ \bibinfo {pages} {025002} (\bibinfo {year}
  {2020})}\BibitemShut {NoStop}%
\bibitem [{\citenamefont {Drautz}(2019{\natexlab{a}})}]{drautz2019atomic}%
  \BibitemOpen
  \bibfield  {author} {\bibinfo {author} {\bibfnamefont {R.}~\bibnamefont
  {Drautz}},\ }\bibfield  {title} {\bibinfo {title} {Atomic cluster expansion
  for accurate and transferable interatomic potentials},\ }\href@noop {}
  {\bibfield  {journal} {\bibinfo  {journal} {Physical Review B}\ }\textbf
  {\bibinfo {volume} {99}},\ \bibinfo {pages} {014104} (\bibinfo {year}
  {2019}{\natexlab{a}})}\BibitemShut {NoStop}%
\bibitem [{\citenamefont {Bochkarev}\ \emph {et~al.}(2022)\citenamefont
  {Bochkarev}, \citenamefont {Lysogorskiy}, \citenamefont {Menon},
  \citenamefont {Qamar}, \citenamefont {Mrovec},\ and\ \citenamefont
  {Drautz}}]{Bochkarev2022}%
  \BibitemOpen
  \bibfield  {author} {\bibinfo {author} {\bibfnamefont {A.}~\bibnamefont
  {Bochkarev}}, \bibinfo {author} {\bibfnamefont {Y.}~\bibnamefont
  {Lysogorskiy}}, \bibinfo {author} {\bibfnamefont {S.}~\bibnamefont {Menon}},
  \bibinfo {author} {\bibfnamefont {M.}~\bibnamefont {Qamar}}, \bibinfo
  {author} {\bibfnamefont {M.}~\bibnamefont {Mrovec}},\ and\ \bibinfo {author}
  {\bibfnamefont {R.}~\bibnamefont {Drautz}},\ }\bibfield  {title} {\bibinfo
  {title} {Efficient parametrization of the atomic cluster expansion},\ }\href
  {https://doi.org/10.1103/PhysRevMaterials.6.013804} {\bibfield  {journal}
  {\bibinfo  {journal} {Phys. Rev. Mater.}\ }\textbf {\bibinfo {volume} {6}},\
  \bibinfo {pages} {013804} (\bibinfo {year} {2022})}\BibitemShut {NoStop}%
\bibitem [{\citenamefont {Yue}\ \emph {et~al.}(2021)\citenamefont {Yue},
  \citenamefont {Muniz}, \citenamefont {Calegari~Andrade}, \citenamefont
  {Zhang}, \citenamefont {Car},\ and\ \citenamefont
  {Panagiotopoulos}}]{yue2021short}%
  \BibitemOpen
  \bibfield  {author} {\bibinfo {author} {\bibfnamefont {S.}~\bibnamefont
  {Yue}}, \bibinfo {author} {\bibfnamefont {M.~C.}\ \bibnamefont {Muniz}},
  \bibinfo {author} {\bibfnamefont {M.~F.}\ \bibnamefont {Calegari~Andrade}},
  \bibinfo {author} {\bibfnamefont {L.}~\bibnamefont {Zhang}}, \bibinfo
  {author} {\bibfnamefont {R.}~\bibnamefont {Car}},\ and\ \bibinfo {author}
  {\bibfnamefont {A.~Z.}\ \bibnamefont {Panagiotopoulos}},\ }\bibfield  {title}
  {\bibinfo {title} {When do short-range atomistic machine-learning models fall
  short?},\ }\href@noop {} {\bibfield  {journal} {\bibinfo  {journal} {The
  Journal of Chemical Physics}\ }\textbf {\bibinfo {volume} {154}} (\bibinfo
  {year} {2021})}\BibitemShut {NoStop}%
\bibitem [{\citenamefont {Grisafi}\ and\ \citenamefont
  {Ceriotti}(2019)}]{grisafi2019incorporating}%
  \BibitemOpen
  \bibfield  {author} {\bibinfo {author} {\bibfnamefont {A.}~\bibnamefont
  {Grisafi}}\ and\ \bibinfo {author} {\bibfnamefont {M.}~\bibnamefont
  {Ceriotti}},\ }\bibfield  {title} {\bibinfo {title} {Incorporating long-range
  physics in atomic-scale machine learning},\ }\href@noop {} {\bibfield
  {journal} {\bibinfo  {journal} {The Journal of chemical physics}\ }\textbf
  {\bibinfo {volume} {151}} (\bibinfo {year} {2019})}\BibitemShut {NoStop}%
\bibitem [{\citenamefont {Unke}\ and\ \citenamefont
  {Meuwly}(2019)}]{unke2019physnet}%
  \BibitemOpen
  \bibfield  {author} {\bibinfo {author} {\bibfnamefont {O.~T.}\ \bibnamefont
  {Unke}}\ and\ \bibinfo {author} {\bibfnamefont {M.}~\bibnamefont {Meuwly}},\
  }\bibfield  {title} {\bibinfo {title} {Physnet: A neural network for
  predicting energies, forces, dipole moments, and partial charges},\
  }\href@noop {} {\bibfield  {journal} {\bibinfo  {journal} {Journal of
  chemical theory and computation}\ }\textbf {\bibinfo {volume} {15}},\
  \bibinfo {pages} {3678} (\bibinfo {year} {2019})}\BibitemShut {NoStop}%
\bibitem [{\citenamefont {Ko}\ \emph {et~al.}(2021)\citenamefont {Ko},
  \citenamefont {Finkler}, \citenamefont {Goedecker},\ and\ \citenamefont
  {Behler}}]{ko2021fourth}%
  \BibitemOpen
  \bibfield  {author} {\bibinfo {author} {\bibfnamefont {T.~W.}\ \bibnamefont
  {Ko}}, \bibinfo {author} {\bibfnamefont {J.~A.}\ \bibnamefont {Finkler}},
  \bibinfo {author} {\bibfnamefont {S.}~\bibnamefont {Goedecker}},\ and\
  \bibinfo {author} {\bibfnamefont {J.}~\bibnamefont {Behler}},\ }\bibfield
  {title} {\bibinfo {title} {A fourth-generation high-dimensional neural
  network potential with accurate electrostatics including non-local charge
  transfer},\ }\href@noop {} {\bibfield  {journal} {\bibinfo  {journal} {Nature
  communications}\ }\textbf {\bibinfo {volume} {12}},\ \bibinfo {pages} {398}
  (\bibinfo {year} {2021})}\BibitemShut {NoStop}%
\bibitem [{\citenamefont {Zhang}\ \emph {et~al.}(2022)\citenamefont {Zhang},
  \citenamefont {Wang}, \citenamefont {Muniz}, \citenamefont {Panagiotopoulos},
  \citenamefont {Car} \emph {et~al.}}]{zhang2022deep}%
  \BibitemOpen
  \bibfield  {author} {\bibinfo {author} {\bibfnamefont {L.}~\bibnamefont
  {Zhang}}, \bibinfo {author} {\bibfnamefont {H.}~\bibnamefont {Wang}},
  \bibinfo {author} {\bibfnamefont {M.~C.}\ \bibnamefont {Muniz}}, \bibinfo
  {author} {\bibfnamefont {A.~Z.}\ \bibnamefont {Panagiotopoulos}}, \bibinfo
  {author} {\bibfnamefont {R.}~\bibnamefont {Car}}, \emph {et~al.},\ }\bibfield
   {title} {\bibinfo {title} {A deep potential model with long-range
  electrostatic interactions},\ }\href@noop {} {\bibfield  {journal} {\bibinfo
  {journal} {The Journal of Chemical Physics}\ }\textbf {\bibinfo {volume}
  {156}} (\bibinfo {year} {2022})}\BibitemShut {NoStop}%
\bibitem [{\citenamefont {Rinaldi}\ \emph {et~al.}(2025)\citenamefont
  {Rinaldi}, \citenamefont {Bochkarev}, \citenamefont {Lysogorskiy},\ and\
  \citenamefont {Drautz}}]{Rinaldi2025}%
  \BibitemOpen
  \bibfield  {author} {\bibinfo {author} {\bibfnamefont {M.}~\bibnamefont
  {Rinaldi}}, \bibinfo {author} {\bibfnamefont {A.}~\bibnamefont {Bochkarev}},
  \bibinfo {author} {\bibfnamefont {Y.}~\bibnamefont {Lysogorskiy}},\ and\
  \bibinfo {author} {\bibfnamefont {R.}~\bibnamefont {Drautz}},\ }\bibfield
  {title} {\bibinfo {title} {Charge-constrained atomic cluster expansion},\
  }\href {https://doi.org/10.1103/PhysRevMaterials.9.033802} {\bibfield
  {journal} {\bibinfo  {journal} {Phys. Rev. Mater.}\ }\textbf {\bibinfo
  {volume} {9}},\ \bibinfo {pages} {033802} (\bibinfo {year}
  {2025})}\BibitemShut {NoStop}%
\bibitem [{\citenamefont {Falletta}\ \emph {et~al.}(2025)\citenamefont
  {Falletta}, \citenamefont {Cepellotti}, \citenamefont {Johansson},
  \citenamefont {Tan}, \citenamefont {Descoteaux}, \citenamefont {Musaelian},
  \citenamefont {Owen},\ and\ \citenamefont {Kozinsky}}]{falletta2025unified}%
  \BibitemOpen
  \bibfield  {author} {\bibinfo {author} {\bibfnamefont {S.}~\bibnamefont
  {Falletta}}, \bibinfo {author} {\bibfnamefont {A.}~\bibnamefont
  {Cepellotti}}, \bibinfo {author} {\bibfnamefont {A.}~\bibnamefont
  {Johansson}}, \bibinfo {author} {\bibfnamefont {C.~W.}\ \bibnamefont {Tan}},
  \bibinfo {author} {\bibfnamefont {M.~L.}\ \bibnamefont {Descoteaux}},
  \bibinfo {author} {\bibfnamefont {A.}~\bibnamefont {Musaelian}}, \bibinfo
  {author} {\bibfnamefont {C.~J.}\ \bibnamefont {Owen}},\ and\ \bibinfo
  {author} {\bibfnamefont {B.}~\bibnamefont {Kozinsky}},\ }\bibfield  {title}
  {\bibinfo {title} {Unified differentiable learning of electric response},\
  }\href@noop {} {\bibfield  {journal} {\bibinfo  {journal} {Nature
  Communications}\ }\textbf {\bibinfo {volume} {16}},\ \bibinfo {pages} {4031}
  (\bibinfo {year} {2025})}\BibitemShut {NoStop}%
\bibitem [{\citenamefont {Joll}\ \emph {et~al.}(2024)\citenamefont {Joll},
  \citenamefont {Schienbein}, \citenamefont {Rosso},\ and\ \citenamefont
  {Blumberger}}]{joll2024machine}%
  \BibitemOpen
  \bibfield  {author} {\bibinfo {author} {\bibfnamefont {K.}~\bibnamefont
  {Joll}}, \bibinfo {author} {\bibfnamefont {P.}~\bibnamefont {Schienbein}},
  \bibinfo {author} {\bibfnamefont {K.~M.}\ \bibnamefont {Rosso}},\ and\
  \bibinfo {author} {\bibfnamefont {J.}~\bibnamefont {Blumberger}},\ }\bibfield
   {title} {\bibinfo {title} {Machine learning the electric field response of
  condensed phase systems using perturbed neural network potentials},\
  }\href@noop {} {\bibfield  {journal} {\bibinfo  {journal} {Nature
  Communications}\ }\textbf {\bibinfo {volume} {15}},\ \bibinfo {pages} {8192}
  (\bibinfo {year} {2024})}\BibitemShut {NoStop}%
\bibitem [{\citenamefont {Hirshfeld}(1977)}]{hirshfeld1977bonded}%
  \BibitemOpen
  \bibfield  {author} {\bibinfo {author} {\bibfnamefont {F.~L.}\ \bibnamefont
  {Hirshfeld}},\ }\bibfield  {title} {\bibinfo {title} {Bonded-atom fragments
  for describing molecular charge densities},\ }\href@noop {} {\bibfield
  {journal} {\bibinfo  {journal} {Theoretica chimica acta}\ }\textbf {\bibinfo
  {volume} {44}},\ \bibinfo {pages} {129} (\bibinfo {year} {1977})}\BibitemShut
  {NoStop}%
\bibitem [{\citenamefont {Shaidu}\ \emph {et~al.}(2024)\citenamefont {Shaidu},
  \citenamefont {Pellegrini}, \citenamefont {K{\"u}{\c{c}}{\"u}kbenli},
  \citenamefont {Lot},\ and\ \citenamefont
  {de~Gironcoli}}]{shaidu2024incorporating}%
  \BibitemOpen
  \bibfield  {author} {\bibinfo {author} {\bibfnamefont {Y.}~\bibnamefont
  {Shaidu}}, \bibinfo {author} {\bibfnamefont {F.}~\bibnamefont {Pellegrini}},
  \bibinfo {author} {\bibfnamefont {E.}~\bibnamefont
  {K{\"u}{\c{c}}{\"u}kbenli}}, \bibinfo {author} {\bibfnamefont
  {R.}~\bibnamefont {Lot}},\ and\ \bibinfo {author} {\bibfnamefont
  {S.}~\bibnamefont {de~Gironcoli}},\ }\bibfield  {title} {\bibinfo {title}
  {Incorporating long-range electrostatics in neural network potentials via
  variational charge equilibration from shortsighted ingredients},\ }\href@noop
  {} {\bibfield  {journal} {\bibinfo  {journal} {npj Computational Materials}\
  }\textbf {\bibinfo {volume} {10}},\ \bibinfo {pages} {47} (\bibinfo {year}
  {2024})}\BibitemShut {NoStop}%
\bibitem [{\citenamefont {Kocer}\ \emph {et~al.}(2025)\citenamefont {Kocer},
  \citenamefont {Singraber}, \citenamefont {Finkler}, \citenamefont {Misof},
  \citenamefont {Ko}, \citenamefont {Dellago},\ and\ \citenamefont
  {Behler}}]{kocer2025iterative}%
  \BibitemOpen
  \bibfield  {author} {\bibinfo {author} {\bibfnamefont {E.}~\bibnamefont
  {Kocer}}, \bibinfo {author} {\bibfnamefont {A.}~\bibnamefont {Singraber}},
  \bibinfo {author} {\bibfnamefont {J.~A.}\ \bibnamefont {Finkler}}, \bibinfo
  {author} {\bibfnamefont {P.}~\bibnamefont {Misof}}, \bibinfo {author}
  {\bibfnamefont {T.~W.}\ \bibnamefont {Ko}}, \bibinfo {author} {\bibfnamefont
  {C.}~\bibnamefont {Dellago}},\ and\ \bibinfo {author} {\bibfnamefont
  {J.}~\bibnamefont {Behler}},\ }\bibfield  {title} {\bibinfo {title}
  {Iterative charge equilibration for fourth-generation high-dimensional neural
  network potentials},\ }\href@noop {} {\bibfield  {journal} {\bibinfo
  {journal} {The Journal of Chemical Physics}\ }\textbf {\bibinfo {volume}
  {162}} (\bibinfo {year} {2025})}\BibitemShut {NoStop}%
\bibitem [{\citenamefont {Marzari}\ and\ \citenamefont
  {Vanderbilt}(1997)}]{marzari1997maximally}%
  \BibitemOpen
  \bibfield  {author} {\bibinfo {author} {\bibfnamefont {N.}~\bibnamefont
  {Marzari}}\ and\ \bibinfo {author} {\bibfnamefont {D.}~\bibnamefont
  {Vanderbilt}},\ }\bibfield  {title} {\bibinfo {title} {Maximally localized
  generalized wannier functions for composite energy bands},\ }\href@noop {}
  {\bibfield  {journal} {\bibinfo  {journal} {Physical review B}\ }\textbf
  {\bibinfo {volume} {56}},\ \bibinfo {pages} {12847} (\bibinfo {year}
  {1997})}\BibitemShut {NoStop}%
\bibitem [{\citenamefont {Marzari}\ \emph {et~al.}(2012)\citenamefont
  {Marzari}, \citenamefont {Mostofi}, \citenamefont {Yates}, \citenamefont
  {Souza},\ and\ \citenamefont {Vanderbilt}}]{marzari2012maximally}%
  \BibitemOpen
  \bibfield  {author} {\bibinfo {author} {\bibfnamefont {N.}~\bibnamefont
  {Marzari}}, \bibinfo {author} {\bibfnamefont {A.~A.}\ \bibnamefont
  {Mostofi}}, \bibinfo {author} {\bibfnamefont {J.~R.}\ \bibnamefont {Yates}},
  \bibinfo {author} {\bibfnamefont {I.}~\bibnamefont {Souza}},\ and\ \bibinfo
  {author} {\bibfnamefont {D.}~\bibnamefont {Vanderbilt}},\ }\bibfield  {title}
  {\bibinfo {title} {Maximally localized wannier functions: Theory and
  applications},\ }\href@noop {} {\bibfield  {journal} {\bibinfo  {journal}
  {Reviews of Modern Physics}\ }\textbf {\bibinfo {volume} {84}},\ \bibinfo
  {pages} {1419} (\bibinfo {year} {2012})}\BibitemShut {NoStop}%
\bibitem [{\citenamefont {Gao}\ and\ \citenamefont
  {Remsing}(2022)}]{gao2022self}%
  \BibitemOpen
  \bibfield  {author} {\bibinfo {author} {\bibfnamefont {A.}~\bibnamefont
  {Gao}}\ and\ \bibinfo {author} {\bibfnamefont {R.~C.}\ \bibnamefont
  {Remsing}},\ }\bibfield  {title} {\bibinfo {title} {Self-consistent
  determination of long-range electrostatics in neural network potentials},\
  }\href@noop {} {\bibfield  {journal} {\bibinfo  {journal} {Nature
  communications}\ }\textbf {\bibinfo {volume} {13}},\ \bibinfo {pages} {1572}
  (\bibinfo {year} {2022})}\BibitemShut {NoStop}%
\bibitem [{\citenamefont {Zhang}\ \emph {et~al.}(2024)\citenamefont {Zhang},
  \citenamefont {Calegari~Andrade}, \citenamefont {Goldsmith}, \citenamefont
  {Raman}, \citenamefont {Li}, \citenamefont {Piaggi}, \citenamefont {Wu},
  \citenamefont {Car},\ and\ \citenamefont {Selloni}}]{zhang2024molecular}%
  \BibitemOpen
  \bibfield  {author} {\bibinfo {author} {\bibfnamefont {C.}~\bibnamefont
  {Zhang}}, \bibinfo {author} {\bibfnamefont {M.~F.}\ \bibnamefont
  {Calegari~Andrade}}, \bibinfo {author} {\bibfnamefont {Z.~K.}\ \bibnamefont
  {Goldsmith}}, \bibinfo {author} {\bibfnamefont {A.~S.}\ \bibnamefont
  {Raman}}, \bibinfo {author} {\bibfnamefont {Y.}~\bibnamefont {Li}}, \bibinfo
  {author} {\bibfnamefont {P.~M.}\ \bibnamefont {Piaggi}}, \bibinfo {author}
  {\bibfnamefont {X.}~\bibnamefont {Wu}}, \bibinfo {author} {\bibfnamefont
  {R.}~\bibnamefont {Car}},\ and\ \bibinfo {author} {\bibfnamefont
  {A.}~\bibnamefont {Selloni}},\ }\bibfield  {title} {\bibinfo {title}
  {Molecular-scale insights into the electrical double layer at
  oxide-electrolyte interfaces},\ }\href@noop {} {\bibfield  {journal}
  {\bibinfo  {journal} {Nature Communications}\ }\textbf {\bibinfo {volume}
  {15}},\ \bibinfo {pages} {10270} (\bibinfo {year} {2024})}\BibitemShut
  {NoStop}%
\bibitem [{\citenamefont {Zhu}\ and\ \citenamefont
  {Cheng}(2025)}]{zhu2024machine}%
  \BibitemOpen
  \bibfield  {author} {\bibinfo {author} {\bibfnamefont {J.-X.}\ \bibnamefont
  {Zhu}}\ and\ \bibinfo {author} {\bibfnamefont {J.}~\bibnamefont {Cheng}},\
  }\bibfield  {title} {\bibinfo {title} {Machine learning potential for
  electrochemical interfaces with hybrid representation of dielectric
  response},\ }\href {https://doi.org/10.1103/48ct-3jxm} {\bibfield  {journal}
  {\bibinfo  {journal} {Phys. Rev. Lett.}\ }\textbf {\bibinfo {volume} {135}},\
  \bibinfo {pages} {018003} (\bibinfo {year} {2025})}\BibitemShut {NoStop}%
\bibitem [{\citenamefont {Zhang}\ \emph {et~al.}(2020)\citenamefont {Zhang},
  \citenamefont {Chen}, \citenamefont {Wu}, \citenamefont {Wang}, \citenamefont
  {E},\ and\ \citenamefont {Car}}]{Zhang2020}%
  \BibitemOpen
  \bibfield  {author} {\bibinfo {author} {\bibfnamefont {L.}~\bibnamefont
  {Zhang}}, \bibinfo {author} {\bibfnamefont {M.}~\bibnamefont {Chen}},
  \bibinfo {author} {\bibfnamefont {X.}~\bibnamefont {Wu}}, \bibinfo {author}
  {\bibfnamefont {H.}~\bibnamefont {Wang}}, \bibinfo {author} {\bibfnamefont
  {W.}~\bibnamefont {E}},\ and\ \bibinfo {author} {\bibfnamefont
  {R.}~\bibnamefont {Car}},\ }\bibfield  {title} {\bibinfo {title} {Deep neural
  network for the dielectric response of insulators},\ }\href
  {https://doi.org/10.1103/PhysRevB.102.041121} {\bibfield  {journal} {\bibinfo
   {journal} {Phys. Rev. B}\ }\textbf {\bibinfo {volume} {102}},\ \bibinfo
  {pages} {041121} (\bibinfo {year} {2020})}\BibitemShut {NoStop}%
\bibitem [{\citenamefont {Krishnamoorthy}\ \emph {et~al.}(2021)\citenamefont
  {Krishnamoorthy}, \citenamefont {Nomura}, \citenamefont {Baradwaj},
  \citenamefont {Shimamura}, \citenamefont {Rajak}, \citenamefont {Mishra},
  \citenamefont {Fukushima}, \citenamefont {Shimojo}, \citenamefont {Kalia},
  \citenamefont {Nakano},\ and\ \citenamefont
  {Vashishta}}]{Krishnamoorthy2021}%
  \BibitemOpen
  \bibfield  {author} {\bibinfo {author} {\bibfnamefont {A.}~\bibnamefont
  {Krishnamoorthy}}, \bibinfo {author} {\bibfnamefont {K.-i.}\ \bibnamefont
  {Nomura}}, \bibinfo {author} {\bibfnamefont {N.}~\bibnamefont {Baradwaj}},
  \bibinfo {author} {\bibfnamefont {K.}~\bibnamefont {Shimamura}}, \bibinfo
  {author} {\bibfnamefont {P.}~\bibnamefont {Rajak}}, \bibinfo {author}
  {\bibfnamefont {A.}~\bibnamefont {Mishra}}, \bibinfo {author} {\bibfnamefont
  {S.}~\bibnamefont {Fukushima}}, \bibinfo {author} {\bibfnamefont
  {F.}~\bibnamefont {Shimojo}}, \bibinfo {author} {\bibfnamefont
  {R.}~\bibnamefont {Kalia}}, \bibinfo {author} {\bibfnamefont
  {A.}~\bibnamefont {Nakano}},\ and\ \bibinfo {author} {\bibfnamefont
  {P.}~\bibnamefont {Vashishta}},\ }\bibfield  {title} {\bibinfo {title}
  {Dielectric constant of liquid water determined with neural network quantum
  molecular dynamics},\ }\href {https://doi.org/10.1103/PhysRevLett.126.216403}
  {\bibfield  {journal} {\bibinfo  {journal} {Phys. Rev. Lett.}\ }\textbf
  {\bibinfo {volume} {126}},\ \bibinfo {pages} {216403} (\bibinfo {year}
  {2021})}\BibitemShut {NoStop}%
\bibitem [{\citenamefont {Han}\ \emph {et~al.}(2023)\citenamefont {Han},
  \citenamefont {Isborn},\ and\ \citenamefont {Shi}}]{Han2023}%
  \BibitemOpen
  \bibfield  {author} {\bibinfo {author} {\bibfnamefont {B.}~\bibnamefont
  {Han}}, \bibinfo {author} {\bibfnamefont {C.~M.}\ \bibnamefont {Isborn}},\
  and\ \bibinfo {author} {\bibfnamefont {L.}~\bibnamefont {Shi}},\ }\bibfield
  {title} {\bibinfo {title} {Incorporating polarization and charge transfer
  into a point-charge model for water using machine learning},\ }\href
  {https://doi.org/10.1021/acs.jpclett.3c00036} {\bibfield  {journal} {\bibinfo
   {journal} {The Journal of Physical Chemistry Letters}\ }\textbf {\bibinfo
  {volume} {14}},\ \bibinfo {pages} {3869} (\bibinfo {year}
  {2023})}\BibitemShut {NoStop}%
\bibitem [{\citenamefont {Liang}\ and\ \citenamefont
  {Yang}(2025)}]{liang2025polarizable}%
  \BibitemOpen
  \bibfield  {author} {\bibinfo {author} {\bibfnamefont {Q.}~\bibnamefont
  {Liang}}\ and\ \bibinfo {author} {\bibfnamefont {J.}~\bibnamefont {Yang}},\
  }\bibfield  {title} {\bibinfo {title} {Polarizable water model with ab initio
  neural network dynamic charges and spontaneous charge transfer},\ }\href@noop
  {} {\bibfield  {journal} {\bibinfo  {journal} {Journal of Chemical Theory and
  Computation}\ }\textbf {\bibinfo {volume} {21}},\ \bibinfo {pages} {3360}
  (\bibinfo {year} {2025})}\BibitemShut {NoStop}%
\bibitem [{\citenamefont {Surendralal}\ \emph {et~al.}(2018)\citenamefont
  {Surendralal}, \citenamefont {Todorova}, \citenamefont {Finnis},\ and\
  \citenamefont {Neugebauer}}]{Surendralal2018}%
  \BibitemOpen
  \bibfield  {author} {\bibinfo {author} {\bibfnamefont {S.}~\bibnamefont
  {Surendralal}}, \bibinfo {author} {\bibfnamefont {M.}~\bibnamefont
  {Todorova}}, \bibinfo {author} {\bibfnamefont {M.~W.}\ \bibnamefont
  {Finnis}},\ and\ \bibinfo {author} {\bibfnamefont {J.}~\bibnamefont
  {Neugebauer}},\ }\bibfield  {title} {\bibinfo {title} {First-principles
  approach to model electrochemical reactions: Understanding the fundamental
  mechanisms behind mg corrosion},\ }\href
  {https://doi.org/10.1103/PhysRevLett.120.246801} {\bibfield  {journal}
  {\bibinfo  {journal} {Phys. Rev. Lett.}\ }\textbf {\bibinfo {volume} {120}},\
  \bibinfo {pages} {246801} (\bibinfo {year} {2018})}\BibitemShut {NoStop}%
\bibitem [{\citenamefont {Dei\ss{}enbeck}\ \emph {et~al.}(2021)\citenamefont
  {Dei\ss{}enbeck}, \citenamefont {Freysoldt}, \citenamefont {Todorova},
  \citenamefont {Neugebauer},\ and\ \citenamefont
  {Wippermann}}]{Deissenbeck2021}%
  \BibitemOpen
  \bibfield  {author} {\bibinfo {author} {\bibfnamefont {F.}~\bibnamefont
  {Dei\ss{}enbeck}}, \bibinfo {author} {\bibfnamefont {C.}~\bibnamefont
  {Freysoldt}}, \bibinfo {author} {\bibfnamefont {M.}~\bibnamefont {Todorova}},
  \bibinfo {author} {\bibfnamefont {J.}~\bibnamefont {Neugebauer}},\ and\
  \bibinfo {author} {\bibfnamefont {S.}~\bibnamefont {Wippermann}},\ }\bibfield
   {title} {\bibinfo {title} {Dielectric properties of nanoconfined water: A
  canonical thermopotentiostat approach},\ }\href
  {https://doi.org/10.1103/PhysRevLett.126.136803} {\bibfield  {journal}
  {\bibinfo  {journal} {Phys. Rev. Lett.}\ }\textbf {\bibinfo {volume} {126}},\
  \bibinfo {pages} {136803} (\bibinfo {year} {2021})}\BibitemShut {NoStop}%
\bibitem [{\citenamefont {Neugebauer}\ and\ \citenamefont
  {Scheffler}(1992)}]{Neugebauer1992}%
  \BibitemOpen
  \bibfield  {author} {\bibinfo {author} {\bibfnamefont {J.}~\bibnamefont
  {Neugebauer}}\ and\ \bibinfo {author} {\bibfnamefont {M.}~\bibnamefont
  {Scheffler}},\ }\bibfield  {title} {\bibinfo {title} {Adsorbate-substrate and
  adsorbate-adsorbate interactions of na and k adlayers on al(111)},\ }\href
  {https://doi.org/10.1103/PhysRevB.46.16067} {\bibfield  {journal} {\bibinfo
  {journal} {Phys. Rev. B}\ }\textbf {\bibinfo {volume} {46}},\ \bibinfo
  {pages} {16067} (\bibinfo {year} {1992})}\BibitemShut {NoStop}%
\bibitem [{\citenamefont {Kresse}\ and\ \citenamefont {Hafner}(1993)}]{vasp1}%
  \BibitemOpen
  \bibfield  {author} {\bibinfo {author} {\bibfnamefont {G.}~\bibnamefont
  {Kresse}}\ and\ \bibinfo {author} {\bibfnamefont {J.}~\bibnamefont
  {Hafner}},\ }\bibfield  {title} {\bibinfo {title} {Ab initio molecular
  dynamics for liquid metals},\ }\href
  {https://doi.org/10.1103/PhysRevB.47.558} {\bibfield  {journal} {\bibinfo
  {journal} {Phys. Rev. B}\ }\textbf {\bibinfo {volume} {47}},\ \bibinfo
  {pages} {558} (\bibinfo {year} {1993})}\BibitemShut {NoStop}%
\bibitem [{\citenamefont {Kresse}\ and\ \citenamefont
  {Furthm\"uller}(1996)}]{vasp2}%
  \BibitemOpen
  \bibfield  {author} {\bibinfo {author} {\bibfnamefont {G.}~\bibnamefont
  {Kresse}}\ and\ \bibinfo {author} {\bibfnamefont {J.}~\bibnamefont
  {Furthm\"uller}},\ }\bibfield  {title} {\bibinfo {title} {Efficient iterative
  schemes for ab initio total-energy calculations using a plane-wave basis
  set},\ }\href {https://doi.org/10.1103/PhysRevB.54.11169} {\bibfield
  {journal} {\bibinfo  {journal} {Phys. Rev. B}\ }\textbf {\bibinfo {volume}
  {54}},\ \bibinfo {pages} {11169} (\bibinfo {year} {1996})}\BibitemShut
  {NoStop}%
\bibitem [{\citenamefont {Mostofi}\ \emph {et~al.}(2008)\citenamefont
  {Mostofi}, \citenamefont {Yates}, \citenamefont {Lee}, \citenamefont {Souza},
  \citenamefont {Vanderbilt},\ and\ \citenamefont
  {Marzari}}]{mostofi2008wannier90}%
  \BibitemOpen
  \bibfield  {author} {\bibinfo {author} {\bibfnamefont {A.~A.}\ \bibnamefont
  {Mostofi}}, \bibinfo {author} {\bibfnamefont {J.~R.}\ \bibnamefont {Yates}},
  \bibinfo {author} {\bibfnamefont {Y.-S.}\ \bibnamefont {Lee}}, \bibinfo
  {author} {\bibfnamefont {I.}~\bibnamefont {Souza}}, \bibinfo {author}
  {\bibfnamefont {D.}~\bibnamefont {Vanderbilt}},\ and\ \bibinfo {author}
  {\bibfnamefont {N.}~\bibnamefont {Marzari}},\ }\bibfield  {title} {\bibinfo
  {title} {wannier90: A tool for obtaining maximally-localised wannier
  functions},\ }\href@noop {} {\bibfield  {journal} {\bibinfo  {journal}
  {Computer physics communications}\ }\textbf {\bibinfo {volume} {178}},\
  \bibinfo {pages} {685} (\bibinfo {year} {2008})}\BibitemShut {NoStop}%
\bibitem [{\citenamefont {Bader}\ and\ \citenamefont
  {Matta}(2001)}]{bader2001properties}%
  \BibitemOpen
  \bibfield  {author} {\bibinfo {author} {\bibfnamefont {R.~F.}\ \bibnamefont
  {Bader}}\ and\ \bibinfo {author} {\bibfnamefont {C.~F.}\ \bibnamefont
  {Matta}},\ }\bibfield  {title} {\bibinfo {title} {Properties of atoms in
  crystals: dielectric polarization},\ }\href@noop {} {\bibfield  {journal}
  {\bibinfo  {journal} {International Journal of Quantum Chemistry}\ }\textbf
  {\bibinfo {volume} {85}},\ \bibinfo {pages} {592} (\bibinfo {year}
  {2001})}\BibitemShut {NoStop}%
\bibitem [{\citenamefont {Henkelman}\ \emph {et~al.}(2006)\citenamefont
  {Henkelman}, \citenamefont {Arnaldsson},\ and\ \citenamefont
  {J{\'o}nsson}}]{henkelman2006fast}%
  \BibitemOpen
  \bibfield  {author} {\bibinfo {author} {\bibfnamefont {G.}~\bibnamefont
  {Henkelman}}, \bibinfo {author} {\bibfnamefont {A.}~\bibnamefont
  {Arnaldsson}},\ and\ \bibinfo {author} {\bibfnamefont {H.}~\bibnamefont
  {J{\'o}nsson}},\ }\bibfield  {title} {\bibinfo {title} {A fast and robust
  algorithm for bader decomposition of charge density},\ }\href@noop {}
  {\bibfield  {journal} {\bibinfo  {journal} {Computational Materials Science}\
  }\textbf {\bibinfo {volume} {36}},\ \bibinfo {pages} {354} (\bibinfo {year}
  {2006})}\BibitemShut {NoStop}%
\bibitem [{SM()}]{SM}%
  \BibitemOpen
  \href@noop {} {\bibinfo  {journal} {See Supplemental Material for
  computational details}\ }\BibitemShut {NoStop}%
\bibitem [{\citenamefont {Drautz}(2019{\natexlab{b}})}]{Drautz2019}%
  \BibitemOpen
\bibfield  {journal} {  }\bibfield  {author} {\bibinfo {author} {\bibfnamefont
  {R.}~\bibnamefont {Drautz}},\ }\bibfield  {title} {\bibinfo {title} {Atomic
  cluster expansion for accurate and transferable interatomic potentials},\
  }\href {https://doi.org/10.1103/PhysRevB.99.014104} {\bibfield  {journal}
  {\bibinfo  {journal} {Phys. Rev. B}\ }\textbf {\bibinfo {volume} {99}},\
  \bibinfo {pages} {014104} (\bibinfo {year} {2019}{\natexlab{b}})}\BibitemShut
  {NoStop}%
\bibitem [{\citenamefont {Lysogorskiy}\ \emph {et~al.}(2021)\citenamefont
  {Lysogorskiy}, \citenamefont {Oord}, \citenamefont {Bochkarev}, \citenamefont
  {Menon}, \citenamefont {Rinaldi}, \citenamefont {Hammerschmidt},
  \citenamefont {Mrovec}, \citenamefont {Thompson}, \citenamefont {Cs{\'a}nyi},
  \citenamefont {Ortner},\ and\ \citenamefont
  {Drautz}}]{lysogorskiy2021performant}%
  \BibitemOpen
  \bibfield  {author} {\bibinfo {author} {\bibfnamefont {Y.}~\bibnamefont
  {Lysogorskiy}}, \bibinfo {author} {\bibfnamefont {C.~v.~d.}\ \bibnamefont
  {Oord}}, \bibinfo {author} {\bibfnamefont {A.}~\bibnamefont {Bochkarev}},
  \bibinfo {author} {\bibfnamefont {S.}~\bibnamefont {Menon}}, \bibinfo
  {author} {\bibfnamefont {M.}~\bibnamefont {Rinaldi}}, \bibinfo {author}
  {\bibfnamefont {T.}~\bibnamefont {Hammerschmidt}}, \bibinfo {author}
  {\bibfnamefont {M.}~\bibnamefont {Mrovec}}, \bibinfo {author} {\bibfnamefont
  {A.}~\bibnamefont {Thompson}}, \bibinfo {author} {\bibfnamefont
  {G.}~\bibnamefont {Cs{\'a}nyi}}, \bibinfo {author} {\bibfnamefont
  {C.}~\bibnamefont {Ortner}},\ and\ \bibinfo {author} {\bibfnamefont
  {R.}~\bibnamefont {Drautz}},\ }\bibfield  {title} {\bibinfo {title}
  {Performant implementation of the atomic cluster expansion (pace) and
  application to copper and silicon},\ }\href
  {https://doi.org/https://doi.org/10.1038/s41524-021-00559-9} {\bibfield
  {journal} {\bibinfo  {journal} {npj Computational Materials}\ }\textbf
  {\bibinfo {volume} {7}},\ \bibinfo {pages} {97} (\bibinfo {year}
  {2021})}\BibitemShut {NoStop}%
\bibitem [{\citenamefont {Lysogorskiy}\ \emph {et~al.}(2023)\citenamefont
  {Lysogorskiy}, \citenamefont {Bochkarev}, \citenamefont {Mrovec},\ and\
  \citenamefont {Drautz}}]{Lysogorskiy2023}%
  \BibitemOpen
  \bibfield  {author} {\bibinfo {author} {\bibfnamefont {Y.}~\bibnamefont
  {Lysogorskiy}}, \bibinfo {author} {\bibfnamefont {A.}~\bibnamefont
  {Bochkarev}}, \bibinfo {author} {\bibfnamefont {M.}~\bibnamefont {Mrovec}},\
  and\ \bibinfo {author} {\bibfnamefont {R.}~\bibnamefont {Drautz}},\
  }\bibfield  {title} {\bibinfo {title} {Active learning strategies for atomic
  cluster expansion models},\ }\href
  {https://doi.org/10.1103/PhysRevMaterials.7.043801} {\bibfield  {journal}
  {\bibinfo  {journal} {Phys. Rev. Mater.}\ }\textbf {\bibinfo {volume} {7}},\
  \bibinfo {pages} {043801} (\bibinfo {year} {2023})}\BibitemShut {NoStop}%
\bibitem [{\citenamefont {Surendralal}(2020)}]{PhD_Surendralal}%
  \BibitemOpen
  \bibfield  {author} {\bibinfo {author} {\bibfnamefont {S.}~\bibnamefont
  {Surendralal}},\ }\href
  {https://hss-opus.ub.ruhr-uni-bochum.de/opus4/frontdoor/index/index/year/2020/docId/7010}
  {\emph {\bibinfo {title} {Development of an {it ab initio} computational
  potentiostat and its application to the study of Mg corrosion}}}\ (\bibinfo
  {publisher} {PhD Thesis, Ruhr-Universit\"at Bochum, page 71},\ \bibinfo
  {year} {2020})\BibitemShut {NoStop}%
\bibitem [{\citenamefont {Beetz}(1866)}]{beetz1866xxxiv}%
  \BibitemOpen
  \bibfield  {author} {\bibinfo {author} {\bibfnamefont {W.}~\bibnamefont
  {Beetz}},\ }\bibfield  {title} {\bibinfo {title} {Xxxiv. on the development
  of hydrogen from the anode},\ }\href@noop {} {\bibfield  {journal} {\bibinfo
  {journal} {The London, Edinburgh, and Dublin Philosophical Magazine and
  Journal of Science}\ }\textbf {\bibinfo {volume} {32}},\ \bibinfo {pages}
  {269} (\bibinfo {year} {1866})}\BibitemShut {NoStop}%
\bibitem [{\citenamefont {Song}\ and\ \citenamefont
  {Atrens}(1999)}]{song1999corrosion}%
  \BibitemOpen
  \bibfield  {author} {\bibinfo {author} {\bibfnamefont {G.~L.}\ \bibnamefont
  {Song}}\ and\ \bibinfo {author} {\bibfnamefont {A.}~\bibnamefont {Atrens}},\
  }\bibfield  {title} {\bibinfo {title} {Corrosion mechanisms of magnesium
  alloys},\ }\href@noop {} {\bibfield  {journal} {\bibinfo  {journal} {Advanced
  engineering materials}\ }\textbf {\bibinfo {volume} {1}},\ \bibinfo {pages}
  {11} (\bibinfo {year} {1999})}\BibitemShut {NoStop}%
\bibitem [{\citenamefont {Yuwono}\ \emph {et~al.}(2019)\citenamefont {Yuwono},
  \citenamefont {Taylor}, \citenamefont {Frankel}, \citenamefont {Birbilis},\
  and\ \citenamefont {Fajardo}}]{yuwono2019understanding}%
  \BibitemOpen
  \bibfield  {author} {\bibinfo {author} {\bibfnamefont {J.~A.}\ \bibnamefont
  {Yuwono}}, \bibinfo {author} {\bibfnamefont {C.~D.}\ \bibnamefont {Taylor}},
  \bibinfo {author} {\bibfnamefont {G.~S.}\ \bibnamefont {Frankel}}, \bibinfo
  {author} {\bibfnamefont {N.}~\bibnamefont {Birbilis}},\ and\ \bibinfo
  {author} {\bibfnamefont {S.}~\bibnamefont {Fajardo}},\ }\bibfield  {title}
  {\bibinfo {title} {Understanding the enhanced rates of hydrogen evolution on
  dissolving magnesium},\ }\href@noop {} {\bibfield  {journal} {\bibinfo
  {journal} {Electrochemistry Communications}\ }\textbf {\bibinfo {volume}
  {104}},\ \bibinfo {pages} {106482} (\bibinfo {year} {2019})}\BibitemShut
  {NoStop}%
\bibitem [{\citenamefont {Dei{\ss}enbeck}\ \emph {et~al.}(2024)\citenamefont
  {Dei{\ss}enbeck}, \citenamefont {Surendralal}, \citenamefont {Todorova},
  \citenamefont {Wippermann},\ and\ \citenamefont
  {Neugebauer}}]{deissenbeck2024revealing}%
  \BibitemOpen
  \bibfield  {author} {\bibinfo {author} {\bibfnamefont {F.}~\bibnamefont
  {Dei{\ss}enbeck}}, \bibinfo {author} {\bibfnamefont {S.}~\bibnamefont
  {Surendralal}}, \bibinfo {author} {\bibfnamefont {M.}~\bibnamefont
  {Todorova}}, \bibinfo {author} {\bibfnamefont {S.}~\bibnamefont
  {Wippermann}},\ and\ \bibinfo {author} {\bibfnamefont {J.}~\bibnamefont
  {Neugebauer}},\ }\bibfield  {title} {\bibinfo {title} {Revealing the reaction
  pathway of anodic hydrogen evolution at magnesium surfaces in aqueous
  electrolytes},\ }\href@noop {} {\bibfield  {journal} {\bibinfo  {journal}
  {Journal of the American Chemical Society}\ }\textbf {\bibinfo {volume}
  {146}},\ \bibinfo {pages} {30314} (\bibinfo {year} {2024})}\BibitemShut
  {NoStop}%
\bibitem [{\citenamefont {Li}\ \emph {et~al.}(2024)\citenamefont {Li},
  \citenamefont {Harrison},\ and\ \citenamefont
  {Horsfield}}]{li2024uncovering}%
  \BibitemOpen
  \bibfield  {author} {\bibinfo {author} {\bibfnamefont {B.}~\bibnamefont
  {Li}}, \bibinfo {author} {\bibfnamefont {N.~M.}\ \bibnamefont {Harrison}},\
  and\ \bibinfo {author} {\bibfnamefont {A.~P.}\ \bibnamefont {Horsfield}},\
  }\bibfield  {title} {\bibinfo {title} {Uncovering the electrochemical
  stability and corrosion reaction pathway of mg (0001) surface: Insight from
  first-principles calculation},\ }\href@noop {} {\bibfield  {journal}
  {\bibinfo  {journal} {Corrosion Science}\ }\textbf {\bibinfo {volume}
  {241}},\ \bibinfo {pages} {112524} (\bibinfo {year} {2024})}\BibitemShut
  {NoStop}%
\end{thebibliography}%
\end{document}